\documentclass[12pt]{article}
\usepackage{amsmath,amssymb,amsfonts,epsf}
\def\ci{\mathrm{i}}
\def\ud{\mathrm{d}}
\def\Tr{\mathrm{Tr}}

\newcommand{\expp}{\mathrm{e}}
\renewcommand{\Re}{\operatorname{Re}}
\renewcommand{\Im}{\operatorname{Im}}

\newcommand{\q}{\boldsymbol q}
\newcommand{\p}{\boldsymbol p}
\newcommand{\x}{\boldsymbol x}

\begin{document}
~\vspace{2.00cm}

\centerline{\LARGE Wavefunctions, Green's functions and expectation values}
\vspace{0.3cm}
\centerline{\LARGE in terms of spectral determinants}
\vspace{2.0cm}
\centerline{\large 
Martin Sieber$^{1,2}$ \footnote{E-mail: m.sieber@bristol.ac.uk}}
\vspace{0.5cm}

\centerline{$^1$ School of Mathematics, University of Bristol,
Bristol BS8\,1TW, UK}
\centerline{$^2$ Centro Brasileiro de Pesquisas F\'isicas, Rua Dr. Xavier Sigaud 150,
22290-180, Rio de Janeiro, Brasil}

\vspace{4.0cm}
\centerline{\bf Abstract}
\vspace{0.5cm}

We derive semiclassical approximations for wavefunctions, Green's functions
and expectation values for classically chaotic quantum systems. Our method
consists of applying singular and regular perturbations to quantum
Hamiltonians. The wavefunctions, Green's functions and expectation values
of the unperturbed Hamiltonian are expressed in terms of the spectral
determinant of the perturbed Hamiltonian. Semiclassical resummation methods
for spectral determinants are applied and yield approximations in terms
of a finite number of classical trajectories. The final
formulas have a simple form. In contrast to Poincar\'e surface of
section methods, the resummation is done in terms of the periods of the
trajectories.

\vspace{3.5cm}

\noindent PACS numbers: \\
\noindent 03.65.Sq ~ Semiclassical theories and applications. \\
\noindent 05.45.Mt ~ Semiclassical chaos (``quantum chaos'').

\clearpage

\section{Introduction}

Semiclassical methods have a central role in the study
of classically chaotic quantum systems, because they
allow to explain properties of quantum systems on the
basis of the underlying classical dynamics. A prominent
example is the Gutzwiller trace formula that approximates
the density of states by a sum over classical periodic orbits.
Among its many applications are the explanation of periodic
orbit signatures in quantum spectra and the semiclassical
evaluation of spectral statistics \cite{Gut90,Sto99}.
For a semiclassical calculation of individual quantum energy
levels, however, it is more efficient to use formulations
in terms of spectral determinants instead of the density
of states. Resummation methods for semiclassical spectral
determinants allow to express contributions of long orbits
in terms of short ones and yield approximations in terms of
a finite number of orbits \cite{BK90,BK92,Bog92,DS92}. As a
consequence, a significantly smaller number of periodic
orbits is needed. Numerical tests show that these resummed
semiclassical sums can be remarkably accurate approximations
to quantum spectral determinants \cite{KS94}.

Similar methods for wavefunctions and related quantities
have been developed in several articles. Bogomolny derived
a semiclassical theory for averages of wavefunctions over
small windows of energy and position, starting from the
semiclassical Green's function \cite{Bog88}.
A corresponding approach in phase space was developed
by Berry \cite{Ber89}. One main motivation behind these
theories was to explain the scar phenomenon \cite{Hel84}.
They apply, however, to averages over several wavefunctions
within the considered energy window and do not involve
resummations. The first semiclassical formula for single
wavefunctions was obtained by Agam and Fishman \cite{AF93}.
Their formula is based on a semiclassical resummation and
was used to accurately predict the occurrence of scars in
wavefunctions \cite{AF94}.
In subsequent articles Fredholm methods for integral
equations on quantum Poincar\'e surfaces of section were
applied to obtain simpler expressions and put the work of
Agam and Fishman on a more rigorous basis
\cite{GP95a,FGP96,Pra98,SS00,SM01}. In these 
approaches the resummation is done in terms of the
number of intersections of the trajectories with the
Poincar\'e surface instead of their periods. In \cite{AF99}
special properties of tiling billiards on the pseudo-sphere
were used to obtain a resummation in terms of the periods
for these systems.

In this article we use a different approach. We 
apply perturbations to a Hamilton in order to 
express wavefunctions, Green's functions and
expectation values in terms of spectral determinants.
This allows the application of semiclassical
resummation methods for spectral determinants. 
The resulting formulas are simpler
than previous approximations, although they
are equivalent to them in the semiclassical limit.

The article is organized as follows. In section~2
we introduce spectral determinants and their
semiclassical approximations. In section~3 we
discuss singular rank one and rank two
perturbations of Hamiltonians which are applied
in section~4 to obtain resummed semiclassical
approximations for wavefunctions and Green's
functions. Corresponding results for expectation
values are obtained in section~5, and section~6
contains the conclusions.

\section{Spectral determinants}

Spectral determinants are functions whose zeros are given by the
eigenvalues $E_m$ of a quantum Hamiltonian. They are entire 
functions of the energy $E$, real valued for real $E$, and can
be expressed as a product over the quantum energy levels $E_m$
\begin{equation} \label{delta}
\Delta(E) = \prod_{m=1}^\infty A(E,E_m) \, (E - E_m) \; .
\end{equation}
The coefficients $A(E,E_m)$ are non-vanishing functions of $E$ and make the
product convergent. The specific form of these coefficients depends on the
chosen regularization. Two common forms of regularizing spectral determinants, the
Weierstrass and the zeta regularization, are discussed in appendix \ref{app_a}.

As is also discussed in this appendix, the semiclassical approximation
to the spectral determinant can be obtained from the (regularized)
trace of the Green's function and is expressed in terms of the trajectories
of the corresponding classical system \cite{Vor88}
\begin{equation} \label{delbassc}
\Delta(E) \approx B(E) \expp^{- \ci \pi \bar{N}(E)} \zeta(E) \, .
\end{equation}
Here $B(E)$ is a function which is real for real $E$, see appendix
\ref{app_a}. It will drop out of the final formulas of this article.
$\bar{N}(E)$ is the mean spectral counting function which originates 
from zero-length trajectories. Its leading order asymptotic term 
is given by the Thomas-Fermi or Weyl estimate
\begin{equation} \label{nbar}
\bar{N}(E) \sim (2 \pi \hbar)^{-D} \int \ud^D q \, \ud^D p \; 
\theta(E - H(\q,\p)) \quad \text{as} \quad \hbar \rightarrow 0 \, ,
\end{equation}
where $\theta(E)$ denotes the Heaviside theta function and $D$ is
the dimension.

The function $\zeta(E)$ receives contributions
from the classical periodic orbits. In the case of chaotic systems 
it can be expressed as an Euler product over the periodic orbits. 
Alternatively, the product can be expanded and expressed as a sum
over composite (or pseudo) orbits. We discuss the second form
in the following. To each composite orbit, labeled by $n$, corresponds
a finite set of primitive periodic orbits ${\cal P}_n$, and the action of
the composite orbit is a linear combination of the actions of the periodic
orbits ${\cal S}_n = \sum_{p \in {\cal P}_n} m_p S_p$. The coefficients
$m_p$ are positive integers and depend on $n$ though we do not indicate 
this in the notation. More precisely, the index $n$ labels all possible
finite linear combinations of actions of primitive periodic orbits with
positive coefficients. In addition there is a zero-length contribution,
$n=0$, for which ${\cal S}_0=0$. The zeta function is a sum
over all composite orbits in the form
\begin{equation} \label{zetasemi}
\zeta(E) = \sum_{n=0}^\infty C_n \exp\left( \frac{\ci}{\hbar} {\cal S}_n \right) \, ,
\end{equation}
where
\begin{equation} \label{cn}
C_n = \prod_{p \in {\cal P}_n} \frac{(-1)^{m_p} \sigma_p^{m_p (m_p-1)/2}
\exp\left(- i \frac{\pi}{2} m_p \mu_p - \frac{m_p (m_p-1)}{4} u_p \right)}{
\left| \prod_{l=1}^{m_p} \det(M_p^l - 1) \right|^{1/2}} \, , \quad n \neq 0 \, ,
\end{equation}
and $C_0=1$. Here $\mu_p$ is the Maslov index of the primitive periodic orbit
$p$, and $M_p$ is its stability matrix. $M_p$ has eigenvalues $\sigma_p \expp^{u_p}$
where $\sigma_p$ is the sign of the eigenvalues. In (\ref{zetasemi}) with (\ref{cn})
it is assumed that all periodic orbits are unstable. We note that in one-dimensional
systems the corresponding approximation for $\zeta(E)$ consists of a finite number
of terms, see \cite{Vor88}.

Due to the exponential proliferation of periodic orbits in chaotic systems,
the sum in (\ref{zetasemi}) is not absolutely convergent for real $E$. 
Motivated by the Riemann-Siegel formula for the Riemann zeta function,
Berry and Keating derived a resummation of the sum based on a functional
equation of the spectral determinant in $\hbar$, 
$\Delta(E,\hbar)=\Delta(E,-\hbar)$ \cite{BK92}. (A different approach is
based on the fact that $\Delta(E)$ is real if $E$ is real \cite{Kea92}.) 
The derivation uses
Cauchy integral techniques and the semiclassical sum (\ref{zetasemi})
in regions of the complex $\hbar$-plane where it is absolutely convergent.
This resummation makes the sum effectively finite by expressing the
contribution of longer composite orbits in terms of shorter ones. 
The approximation has the form (in the semiclassical regime)
\begin{equation} \label{deltatrunc}
\Delta(E) \approx B(E) \, 2 \Re \expp^{- \ci \pi \bar{N}(E)}
\sum_{n=0}^\infty C_n \exp\left( \frac{\ci}{\hbar} {\cal S}_n \right) 
\theta^{\text{tr}}(T_H/2 -  {\cal T}_n) \; ,
\end{equation}
where ${\cal T}_n = \ud {\cal S}_n / \ud E$ are the periods of the
composite orbits. $\theta^{\text{tr}}(T)$ is a function which smoothly
truncates the sum.
It goes from zero to one in a region with width of order $\hbar^{1-D/2}$
around $T=0$ (for the optimal choice of one free parameter).
$T_H$ is the Heisenberg time $T_H = 2 \pi \hbar \bar{d}(E)
\propto \hbar^{1-D}$, where $\bar{d}(E)=\bar{N}'(E)$ is the mean
density of states. In \cite{BK92} the truncation function is
given by a series of terms whose leading term contains a complementary
error function, and the correction terms are given by higher order derivatives
of the complementary error function and are different from zero only in the
vicinity of $T=0$. 

Resummed spectral determinants form the basis for the semiclassical
approximations in this article. The arguments of Berry and Keating apply
also to the more general expansions of spectral determinants that are
considered in this article. This is because the spectral determinants satisfy
the same functional equation, and the semiclassical expansions have the
same general structure as in (\ref{delbassc}) with (\ref{zetasemi}).
The difference is that not all of the trajectories that form the
composite orbits are periodic.

In the next section we will consider Hamiltonian with singular perturbations.
In this case there are additional contributions from classical trajectories
which are not periodic. They enter the semiclassical approximation by means
of the Green's function $G(\q,\q',E) = \langle \q| (E - H)^{-1} |\q' \rangle$
whose semiclassical approximation is given by \cite{Gut90}
\begin{equation} \label{greensemi}
G(\q,\q',E) \approx \sum_\gamma A_\gamma \expp^{\ci S_\gamma/\hbar} \, , 
\end{equation}
where
\begin{equation} \label{agamma}
A_\gamma = \frac{2 \pi}{(2 \pi \ci \hbar)^{(D+1)/2}}
\sqrt{\left| \frac{1}{\dot{q} \dot{q}'}
{\det}' \left( \frac{\partial^2 S_\gamma}{\partial \q \partial \q'}\right) \right|}
\expp^{- i \nu_\gamma \pi/2} \, .
\end{equation}
Here $\gamma$ labels all classical trajectories at energy $E$ from $\q'$ to $\q$.
$S_\gamma$ is the action and $\nu_\gamma$ the number of conjugate
points along $\gamma$. The prime at the determinant in (\ref{agamma}) indicates
that the determinant involves local coordinates perpendicular to the trajectory.

The approximation (\ref{greensemi}) is not correct in the limit $\q' \rightarrow \q$,
because it does not describe correctly the contribution of those trajectories whose
length goes to zero in this limit. The correct contribution of these short-length
trajectories is given in this limit, for Hamiltonians of the form $H=\p^2/2 m + V(\q)$,
by the free Green's function \cite{Gut67,BM72}
\begin{equation} \label{greenfree}
G^{\text{free}} (\q,\q',E) = \frac{m}{2 i \hbar^2} \left(
\frac{p}{2 \pi \hbar |\q - \q'|}
\right)^{D/2-1}
H^{(1)}_{D/2-1}(p \, |\q-\q'| / \hbar) \; ,
\end{equation}
where $H^{(1)}_{D/2-1}$ denotes a Hankel function of the first
kind \cite{AS64}. The classical momentum $p$ is evaluated at
$(\q+\q')/2$. Formula (\ref{greenfree}) shows the divergence
of the Green's function as $\q' \rightarrow \q$ for $D > 1$.
The divergence is logarithmic for $D=2$ and proportional
to $1/|\q-\q'|$ for $D=3$.

From the imaginary part of expression (\ref{greenfree}) one can obtain
the Thomas-Fermi approximation for the local density of states
\begin{equation} \label{locdens}
d(\q,E) := \sum_{m=1}^\infty | \psi(\q)|^2 \delta(E - E_m)
= - \frac{1}{\pi} \left. \Im G(\q,\q,E + \ci \varepsilon)
\right|_{\varepsilon \rightarrow 0} \; ,
\end{equation}
in the form 
\begin{equation} \label{locdens2}
\bar{d}(\q,E) \approx \frac{2 m \pi^{D/2} p^{D-2}}{(2 \pi \hbar)^D \Gamma(D/2)}
= (2 \pi \hbar)^{-D} \int \ud^D p \; \delta\left(E - \frac{p^2}{2 m} - 
V(\q) \right) \; .
\end{equation}
Integration over $\q$ and $E$ yields (\ref{nbar}) for the
Hamiltonians considered.

For convenience in the next sections we introduce the following notation.
Let $a$ be any quantity which has a semiclassical expansion in terms of
classical trajectories or composite trajectories in the form
$a^{\text{sc}} = \sum_n a_n \expp^{\ci s_n/\hbar}$.
Then we denote the smoothly truncated semiclassical sum by
\begin{equation} \label{notation}
a^{\text{sc,tr}} = \sum_n a_n \expp^{\ci s_n/\hbar} \, 
\theta^{\text{tr}}(T_H/2 - t_n) \, ,
\end{equation}
where sc stands for semiclassical and tr for truncation.
In this notation $\Delta(E) \approx 2 \Re \Delta^{\text{sc,tr}}(E)$,
and the semiclassical energies satisfy the condition
\begin{equation} \label{esc}
2 \Re \Delta^{\text{sc,tr}}(E_m^{\text{sc}}) = 0 \, .
\end{equation}

\section{Singular perturbations} \label{secpoint}

We discuss in this section singular rank one and rank two
perturbations as a means to derive semiclassical approximations
for wavefunctions and Green's functions in the next section.
These kind of perturbations can be defined by the applying
the method of self-adjoint extension.
It is instructive to start with the example of a one-dimensional
delta-function potential, for which the correct results can
be obtained by simpler calculations. 

Consider a quantum Hamiltonian of the form
\begin{equation} \label{hperturb}
H = H_0 + \lambda A \; ,
\end{equation}
where $H_0$ is the unperturbed Hamiltonian and the
perturbation has the form $A = |q \rangle \langle q|$. 
The eigenstates of $H$ are determined by the condition
\begin{equation}
0 = (H_0 - E) \, | \psi \rangle + \lambda | q \rangle \; 
\langle q | \psi \rangle \, .
\end{equation}
Multiplying this equation from the left by $\langle q | \, G_0$,
where $G_0 = (E - H_0)^{-1}$ is the resolvent of the unperturbed
Hamiltonian $H_0$, yields
\begin{equation} \label{eigcond}
0 = (1 - \lambda \langle q | G_0 | q \rangle ) \,
\langle q | \psi \rangle \; .
\end{equation}
The solutions are given by unperturbed energies $E_m$ if the
corresponding eigenstate vanishes at $q$, and by the zeros of
the bracket on the right-hand side.

The resolvent of the perturbed Hamiltonian can be obtained from the
general relation $G = G_0 + G_0 \lambda A G$ from which follows that
\begin{equation} \label{resolvents}
G = \frac{1}{1 - G_0 \lambda | q \rangle \langle q |} G_0 \; .
\end{equation}
The eigenvalues of the Hamiltonian $H$ are given by the poles 
of the resolvent $G$. Hence the new spectral determinant has the form
\begin{equation} \label{detpert}
\Delta(E) = \Delta_0(E) \, \det(1 - \lambda G_0 | q \rangle \langle q |) \; .
\end{equation}
Applying $\det M = \expp^{\Tr \log M}$ yields
\begin{equation} \label{fredholm}
\det(1 - \lambda G_0 | q \rangle \langle q |) 
= \expp^{- \sum_{n=1}^\infty
\frac{\lambda^n}{n} \Tr (G_0 |q \rangle \langle q|)^n}
= \expp^{- \sum_{n=1}^\infty
\frac{\lambda^n}{n} (\langle q | G_0 | q \rangle)^n}
= 1 - \lambda \langle q | G_0 | q \rangle  \, ,
\end{equation}
and the spectral determinant follows as
\begin{equation} \label{detpoint}
\Delta(E) = \Delta_0(E) \, [ 1 - \lambda G_0(q,q,E) ] \, ,
\end{equation}
where $G_0(q,q',E) = \langle q | G_0 | q' \rangle$ is the Green's function.
Note that the poles of the Green's function cancel the zeros of $\Delta_0(E)$
in (\ref{detpoint}), if they are non-degenerate and the corresponding
wavefunction does not vanish at $q$. The zeros of $\Delta(E)$
are given by the zeros of the expression in the bracket, 
plus the zeros of $\Delta_0(E)$ with a degeneracy that is reduced
by one (except if all functions in the eigenspace vanish at $q$).
This agrees with the discussion after equation (\ref{eigcond}).

An alternative presentation of the Green's function can be obtained
by using the relation
\begin{equation}
\frac{1}{1 - G_0 \lambda | q \rangle \langle q |} =
1 + \frac{\lambda \, | q \rangle \langle q | \;  G_0}{
1 - \lambda \, \langle q | G_0 | q \rangle} \, ,
\end{equation}
which can be verified by multiplying both sides with the denominators.
Inserting this into (\ref{resolvents}) results in
\begin{equation} \label{greenpoint}
G(E) = G_0 + G_0 | q \rangle 
\frac{\lambda}{1 - \lambda \langle q | G_0 | q \rangle} 
\langle q | G_0 \, .
\end{equation}
If the fraction is expanded into a geometric series one obtains an expression
that can be interpreted as a sum over diffractive orbits \cite{VWR94}.

It is well known that a delta-function potential is not well-defined in two or
three dimensions. This is reflected by the fact 
that $\langle \q | G_0 | \q \rangle$ is infinite in these cases. The problem
of properly defining a pointlike perturbation for $D=2$ and $D=3$ has a long
history, and the extensive literature about it is reviewed in \cite{AGHH88,AK00}.
There exist several methods for dealing with this problem. The standard
method nowadays is to apply the theory of self-adjoint extension to a Hamiltonian
from whose domain one point has been removed, see e.g. 
\cite{Zor80,AGHH88,Jac91,AK00,Kur04}. The resolvent of the Hamiltonian that
is obtained by the self-adjoint extension
is related by Krein's formula to the
resolvent of the unperturbed Hamiltonian. The resulting expression for the
resolvent of the perturbed Hamiltonian has a very similar form as
(\ref{greenpoint}).  The difference is that the Green's function $G_0(\q,\q,E) =
\langle \q | G_0 | \q \rangle$ is replaced by a regularized expression.
\begin{equation} \label{greenpoint2}
G(E) = G_0 + G_0 |\q \rangle \frac{\lambda}{1 - \lambda G_0^{\text{reg}}(\q,\q,E)}
\langle \q | G_0 \, ,
\end{equation}
and the corresponding expression for the determinant is
\begin{equation} \label{detpointreg}
\Delta_1(E,\lambda,\q) = \Delta_0(E) \, [1 - \lambda G_0^{\text{reg}}(\q,\q,E)] \; .
\end{equation}
The regularization of the Green's function is not unique. Different regularizations
can differ by an arbitrary {\em real} constant \cite{Kur04}. However, all  
different regularizations lead to the same family of self-adjoint extensions
of the Hamiltonian. This can be seen from the fact that a change of the
regularized Green's function in (\ref{greenpoint2}) by an additive constant $c$
can be compensated by redefining the perturbation parameter according to
$\lambda \rightarrow \lambda/(1 + c \lambda)$.

We will regularize the Green's function in the following by subtracting
its divergent part in the limit $\q' \rightarrow \q$.
For Hamiltonians of the form $H_0=\p^2/2 m + V(\q)$
the asymptotic behavior of the Green's function in the limit
$\q' \rightarrow \q$ is given by the free Green's function (\ref{greenfree}),
and we define
\begin{alignat}{2} \label{reg}
G^{\text{reg}}(\q,\q,E) & = \lim_{\q' \rightarrow \q} \left[ G_(\q,\q',E) - 
\frac{m}{\pi \hbar^2} \log(k_0 |\q - \q'|) \right] \quad
& (D = 2) \; ,
\notag \\ 
G^{\text{reg}}(\q,\q,E) & = \lim_{\q' \rightarrow \q} \left[ G_(\q,\q',E) +
\frac{m}{2 \pi \hbar^2 |\q - \q'|} \right] \quad
& (D = 3) \; .
\end{alignat}
The constant $k_0$ in the first line of (\ref{reg}) is arbitrary and is
included for dimensional reasons. It follows from the freedom in the choice
of the regularization that the constant $\lambda$ cannot, in general, be
interpreted as the physical strength of the scatterer. It is interesting
to note that a possible physical interpretation can be obtained by
introducing a new parameter $a$
\begin{equation}
\frac{1}{\lambda} = - \frac{m}{\pi \hbar^2} \log (k_0 a) \quad (D=2) \, , \quad
\frac{1}{\lambda} = \frac{m}{2 \pi \hbar^2 a} \quad (D=3) \; .
\end{equation}
Then formula (\ref{greenpoint2}) is valid for the scattering from a small but
finite scatterer of radius $a$ in the limit that $a$ is much smaller than
the wavelength \cite{ES96}.

The semiclassical approximation for the determinant (\ref{detpointreg}) is
obtained by applying the semiclassical resummation that was discussed in the
last section. It results in
\begin{equation} \label{detpointsc}
\Delta_1(E,\lambda,\q) \approx 2 \Re [\Delta_0(E) \, 
(1 - \lambda G_0^{\text{reg}}(\q,\q,E))]^{\text{sc,tr}} \, .
\end{equation}
This has the following interpretation. $\Delta_0(E)$ and 
$(1 - \lambda G_0^{\text{reg}}(\q,\q,E))$
both have a semiclassical expansion in terms of a sum over classical
trajectories. These two sums are multiplied, and one obtains a sum 
over combined trajectories with actions ${\cal S}_n + S_\gamma$.
These combined trajectories might be viewed as a new form of composite orbits.
The corresponding times along the orbits are ${\cal T}_n + T_\gamma$. The sum 
over the combined trajectories is then smoothly truncated at half the
Heisenberg time.

In the following we will consider also rank two perturbations which correspond
to self-adjoint extensions of a Hamiltonian from whose domain two points are
removed \cite{DG85,AK99,AK00}. They formally correspond to a perturbation by
an operator of the form
$$"\lambda_1 | \q_1 \rangle \langle \q_1 | 
+ \mu | \q_1 \rangle \langle \q_2 |  + \bar{\mu} | \q_2 \rangle \langle \q_1 | 
+ \lambda_2 | \q_2 \rangle \langle \q_2|"$$
where $\lambda_1$ and $\lambda_2$ are real and $\mu=\mu_r + \ci \mu_i$ is complex.
In this case the determinant takes the form
\begin{equation} \label{detgenrank2}
\Delta(E) = \Delta_0(E) \; \det F \; ,
\end{equation}
where $F$ is the following $2 \times 2$ matrix
\begin{equation} \label{amatrix}
F = 1 -
\begin{pmatrix}
G_0^{\text{reg}}(\q_1,\q_1,E) & G_0(\q_1,\q_2,E) \\
G_0(\q_2,\q_1,E) & G_0^{\text{reg}}(\q_2,\q_2,E)
\end{pmatrix} \; 
\begin{pmatrix}
\lambda_1 & \mu \\ \bar{\mu} & \lambda_2
\end{pmatrix} \; .
\end{equation}
We will apply the determinant (\ref{detgenrank2}) in the following for
$\lambda_1 = \lambda_2 = 0$ and denote it by
\begin{equation} \label{detrank2}
\Delta_2(E,\mu,\q_1,\q_2) = \Delta_0(E) \, f(E,\mu,\q_1,\q_2) \; ,
\end{equation}
where the function $f$ follows as
\begin{align} \label{f}
f(E,\mu,\q_1,\q_2)  & = 1 - \mu G_0(\q_2,\q_1,E) - \bar{\mu} G_0(\q_1,\q_2,E) 
\notag \\ & \quad + |\mu|^2 [G_0(\q_1,\q_2,E) \, G_0(\q_2,\q_1,E) 
- G_0^{\text{reg}}(\q_1,\q_1,E) G_0^{\text{reg}}(\q_2,\q_2,E) ] \; .
\end{align}
The determinants $\Delta_2$ and $\Delta_1$ are entire functions of $E$,
because the poles that arise from the Green's functions are 
cancelled by the zeros of $\Delta_0$. 
The corresponding semiclassical approximation for $\Delta_2$
is obtained by inserting the semiclassical expansion for $\Delta_0$
and applying the resummation
\begin{equation} \label{detpointssc}
\Delta_2(E,\mu,\q_1,\q_2) \approx 2 \Re [\Delta_0(E) \, 
f(E,\mu,\q_1,\q_2)]^{\text{sc,tr}} \; ,
\end{equation}
which is to interpreted according to the discussion after
equation (\ref{detpointsc}).

\section{Wavefunctions and Green's functions}
\label{wavesemi}

The effect that a perturbation by a pointlike scatterer has on an energy
level depends on the value of the wavefunction at this point. The 
semiclassical approach in this section is based on this relation.
Consider one non-degenerate eigenvalue $E_m$ of the unperturbed Hamiltonian $H_0$.
The regularized Green's function in the eigenvalue equation (\ref{detpointreg})
has a pole at $E=E_m$ with residue $|\psi_m(\q)|^2$. Hence we find that the
determinant of the perturbed problem, evaluated at the unperturbed energy $E=E_m$,
satisfies
\begin{equation} \label{scstart}
\Delta_1(E_m,\lambda,\q) = - \lambda \lim_{E \rightarrow E_m} \Delta_0(E) 
\frac{|\psi_m(\q)|^2}{E - E_m} = - \lambda \Delta_0'(E_m) |\psi_m(\q)|^2 \; .
\end{equation}
We take a derivative with respect to $\lambda$ on both sides of this equation.
This leads to an expression for the absolute square of the wavefunction as a
quotient of derivatives of the spectral determinant
\begin{equation} \label{psi}
|\psi_m(\q)|^2 = - \left. \frac{\frac{\partial}{\partial \lambda} 
\Delta_1(E,\lambda,\q)}{\frac{\partial}{\partial E} 
\Delta_1(E,\lambda,\q)} \right|_{\substack{\lambda = 0 \\ E = E_m}} \; .
\end{equation}
Inserting the semiclassical approximation (\ref{detpointsc}) results in
\begin{equation} \label{psisc}
|\psi_m(\q)|^2 \approx \frac{\Re [\Delta_0(E_m) \;
G_0(\q,\q,E_m)]^{\text{sc,tr}}}{\Re
\left[\Delta_0^{\prime}(E_m) \right]^{\text{sc,tr}}} \; .
\end{equation}
This is the result for the semiclassical approximation for
the modulus square of the wavefunction. It gives an approximation
in terms of a double sum over composite periodic orbits and 
closed orbits from $\q$ to $\q$ which is cut off smoothly when
the sum of the times along the orbits is half the Heisenberg time.
The energies $E_m$ in (\ref{psisc}) are determined semiclassically
by (\ref{esc}).
 
Let us discuss the difference to previous semiclassical formulas
for the square modulus of the wavefunction. Agam and Fishman 
obtain their approximation by assuming that only closed orbits
which are almost periodic are important \cite{AF93}. They describe
these closed orbits in terms of nearby periodic orbits and perform
a sum over all repetitions of the periodic orbits {\em before} 
applying a resummation to the semiclassical sums. Auslaender
and Fishman \cite{AF99} use the fact that for tiling billiards
on the pseudosphere there exists an exact relation between closed
orbits and periodic orbits. They do not assume that closed orbits
are almost periodic, but perform also the sum over repetitions.
Equation (\ref{psisc}) is simpler than the formulas in these
articles mainly because of this difference in the resummation.
Furthermore, it is not assumed that the contributing closed 
orbits are almost periodic. Other approximations for the square
modulus of the wavefunction were derived by Poincar\'e surface
of section methods \cite{FGP96,KS96,Pra98,SS00}. In these cases
the resummation is based on the semiclassical unitarity of the
transfer operator, or the scattering matrix, and is done in
terms of the number of intersections with the Poincar\'e surface.

We continue with a further discussion of (\ref{psisc}). 
One point to note is that (\ref{psisc}) is independent of the
choice of the regularization of the Green's function.
This is because the real part of the zero-length contribution
to the Green's function is multiplied in (\ref{psisc}) by 
$\Re [\Delta_0(E_m)]^{\text{sc,tr}}$ which
vanishes.
 
There is, however, a non-vanishing contribution to (\ref{psisc})
from the imaginary part of the zero-length contributions to the
Green's function. Using (\ref{locdens}) we find that this
contribution is given by
\begin{equation} \label{zerocontr}
\frac{\Im [ \Delta_0(E_m) ]^{\text{sc,tr}}
}{\Re \left[\Delta_0^{\prime}(E_m) \right]^{\text{sc,tr}}} 
\pi \, \bar{d}(\q,E_m) \; .
\end{equation}
The derivative of the semiclassical determinant, equation
(\ref{delbassc}) with (\ref{nbar}) and (\ref{zetasemi})
consists semiclassically of two terms (the derivative
of $B(E)$ is multiplied by a vanishing term)
\begin{equation} \label{deriv}
\Re [\Delta_0^{\prime}(E_m)]^{\text{sc,tr}} =
\pi \bar{d}(E_m) \, \Im [\Delta_0(E_m) ]^{\text{sc,tr}}
- \frac{1}{\hbar} B(E_m) \Im \left[ \expp^{- \ci \pi \bar{N}(E_m)}
\sum_{n=1}^\infty T_n \, C_n \exp\left( \frac{\ci}{\hbar}
{\cal S}_n \right) \right]^{\text{tr}} .
\end{equation}
Note that $\bar{d}(E) \propto \hbar^{-D}$. If we take only the
first term into account we obtain the following approximation
\begin{equation}
|\psi_m(\q)|^2 \approx \frac{\bar{d}(\q,E_m)}{\bar{d}(E_m)} \; .
\end{equation}
This corresponds to a projection of the classical microcanonical
distribution onto the coordinate space. When integrated over some
small area, this is consistent with the quantum ergodicity theorem
which, roughly speaking, states that almost all matrix elements
of smooth operators tend to the microcanonical average in the
semiclassical limit \cite{Shn74,Col85,Zel87}. 

Equation (\ref{psisc}) can applied for an examination
of scars. If a small smoothing of width $\Delta q$ is applied,
then the initial and final momenta of the contributing closed
orbits differ at most by a value of the order of $\hbar/\Delta q$,
and they can be described in terms of nearby periodic orbits.
Since the $q$-dependence in (\ref{psisc}) is in terms of the
semiclassical Green's function, Bogomolny's analysis of the
structure of wavefunctions near periodic orbits can be applied
without additional smoothing over an energy window.
An effect of the resummation is that the contribution of
a trajectory $\gamma$ is weighted by the semiclassical
expression for the spectral determinant, truncated at
$T_H/2-T_\gamma$. 
The semiclassical criterion for scars of Agam and Fishman
\cite{AF94,AF99} can be obtained by integrating (\ref{psisc})
over some tube around the periodic orbits.

Equation (\ref{psisc}) does not contain any information about the
phase of the wavefunction. It is possible to obtain this information
by a similar approach. For this purpose we consider the determinant
(\ref{detrank2}) for the rank two perturbation. A calculation
similar to (\ref{scstart}) leads to
\begin{equation} \label{psi2}
\psi_m(\q) \overline{\psi_m(\q')} = 
- \left. \frac{\left[\frac{\partial}{\partial \mu_r} +
\ci \frac{\partial}{\partial \mu_i} \right]
\Delta_2(E,\mu,\q,\q')}{2 \frac{\partial}{\partial E} 
\Delta_2(E,\mu,\q,\q')} \right|_{\substack{\mu = 0 \\ E = E_m}} \; .
\end{equation}
We insert the semiclassical approximation (\ref{detpointssc}) and obtain
\begin{equation} \label{psisc2}
\psi_m(\q) \overline{\psi_m(\q')}
\approx \frac{[\Delta_0(E_m) \;
G_0(\q,\q',E_m) + 
\overline{\Delta_0(E_m) \; 
G_0(\q',\q,E_m)}
]^{\text{sc,tr}}}{2 \, \Re
\left[\Delta_0^{\prime}(E_m)
\right]^{\text{sc,tr}}} \; .
\end{equation}
This is a double sum over composite orbits and trajectories
from $\q'$ to $\q$, respectively from $\q$ to $\q'$, which is truncated 
smoothly when the sum of the times along orbits and trajectories is half
the Heisenberg time. Semiclassical approximations for this quantity have
been obtained before by Poincar\'e surface of section methods, see \cite{Pra98}.
(\ref{psisc2}) together with (\ref{psisc}) allows to calculate the
normalized wavefunction semiclassically everywhere.

In the case of an degenerate energy level $E_m$ equations (\ref{psisc}) 
and (\ref{psisc2}) are formulas for $\langle \q | P_m | \q \rangle$
and $\langle \q | P_m | \q' \rangle$, respectively, where $P_m$
is the projection operator onto the corresponding eigenspace.
If the level is $d$-fold degenerate one can choose a basis 
of the eigenspace for which $d-1$ basis states vanish at $\q$.
Hence equations (\ref{psisc}) and (\ref{psisc2}) are formulas
for the remaining state which, in general, does
not vanish at $\q$.

The Green's function can be obtained in a very similar way.
From (\ref{detrank2}) follows that
\begin{equation} \label{delgreen}
\Delta_0(E) \, G_0(\q,\q',E) =  \left. \frac{1}{2} 
\left[ \frac{\partial}{\partial \mu_r} 
+ \ci \frac{\partial}{\partial \mu_i} \right]
\Delta_2(E,\mu,\q,\q') \right|_{\mu = 0} \, .
\end{equation}
We insert the semiclassical approximation for the
determinant (\ref{detpointssc}) and obtain
\begin{equation} \label{greenresum}
G_0(\q,\q',E) \approx \frac{[\Delta_0(E) \;
G_0(E,\q,\q') + 
\overline{\Delta_0(E) \; G_0(E,\q',\q)}
]^{\text{sc,tr}}}{2 \, \Re
\left[\Delta_0(E) \right]^{\text{sc,tr}}} \; .
\end{equation}
As in (\ref{psisc2}) this is a double sum over composite orbits
and open trajectories. Resummed expressions for the Green's function
have been obtained by Poincar\'e section
methods in \cite{GP95a,FGP96,Pra98}, and 
for $G_0^{\text{reg}}(\q,\q,E)$ for tiling
billiards on the pseudosphere in \cite{AF99}.
In appendix \ref{app_b} we discuss an example
of a one-dimensional system for which formula
(\ref{greenresum}), with a sharp cut-off,
yields the exact Green's function.

Given the approximation (\ref{greenresum})
it is straightforward to obtain resummed formulas
for other quantities that are related to the
Green's function. For example, the Wigner transform
of the Green function is
\begin{equation} \label{wigner}
W_0(\x,E) = \int \ud^D q' \; \expp^{- \ci \p \q' / \hbar}
G_0 (\q + \frac{1}{2} \q',\q - \frac{1}{2} \q',E) = h^D
\sum_m \frac{W_m(\x)}{E - E_m} \; ,
\end{equation}
where $\x=(\q,\p)$ and $W_m(\x)$ is the Wigner function
of the $m$-th eigenstate with energy $E_m$. Using
(\ref{greenresum}) and performing the integrals in
(\ref{wigner}) in stationary phase approximation 
results in
\begin{equation} \label{wignerresum}
W_0(\x,E) \approx \frac{[\Delta_0(E) \; W_0(\x,E) + 
\overline{\Delta_0(E) \; W_0(\x,E)}
]^{\text{sc,tr}}}{2 \, \Re
\left[\Delta_0(E) \right]^{\text{sc,tr}}} \; ,
\end{equation}
where the semiclassical approximation to the
Wigner function $W_0$ is given by classical
trajectories that satisfy the midpoint rule
$\x = (\x_f + \x_i)/2$ where $\x_f$ and $\x_i$
are the final and initial points in phase space
\cite{Ber89,Ozo98}. The action terms in
(\ref{wignerresum}) are of the form
${\cal S}_n + S_\gamma - \q' \p$, where
$\q'$ is determined by the stationary phase
condition, but the resummation
is done in terms of ${\cal T}_n + T_\gamma$. 
One may speculate that (\ref{wignerresum}) 
holds also for a resummation in terms 
of $\frac{\ud}{\ud E}[{\cal S}_n + S_\gamma
- \q' \p]$, and that this can be derived,
for example,
by starting with a different perturbation.

The corresponding approximation for the Wigner
functions of the eigenstates is given by
\begin{equation} \label{wignerresum2}
W_m(\x) \approx \frac{[\Delta_0(E_m) \;
W_0(\x,E_m) + 
\overline{\Delta_0(E_m) \; W_0(\x,E_m)}
]^{\text{sc,tr}}}{2 \, h^D \, \Re
\left[\Delta^{\prime}_0(E_m) \right]^{\text{sc,tr}}} \; .
\end{equation}

\section{Expectation values}

In this section we apply a similar approach
to expectation values.
Consider the perturbation of a Hamiltonian by a self-adjoint
operator $A$ in the form
$H = H_0 + \lambda A$, and let $\Delta(E,\lambda)$ be the associated
spectral determinant. As before we denote the determinant of the
unperturbed Hamiltonian by $\Delta_0(E)$. 
In contrast to the previous sections we assume now that $A$ is smooth
so that the spectral determinant is semiclassically given only in terms
of periodic orbits (and not diffractive orbits).
Then, if $E_m(\lambda)$ is a non-degenerate energy level, and
$E_m$ denotes $E_m(0)$, it follows from the form (\ref{delta}) that
\begin{equation}
\left. \frac{\partial}{\partial \lambda} \Delta(E,\lambda)
\right|_{\substack{\lambda = 0 \\ E = E_m}}
= \left. - \frac{\partial E_m(\lambda)}{\partial \lambda}
\right|_{\lambda = 0} \; 
\left. \frac{\Delta_0(E)}{E - E_m} 
\right|_{E \rightarrow E_m}
= - \langle \psi_m | A | \psi_m \rangle \; \Delta_0^{\prime}(E_m) \; ,
\end{equation}
where the second equality follows from first order
perturbation theory. Hence the expectation value of 
the operator $A$ in the state $|\psi_m \rangle$ can
be written as a quotient of two partial derivatives
of the spectral determinant
\begin{equation} \label{delexpt}
\langle \psi_m | A | \psi_m \rangle = - \left.
\frac{\frac{\partial}{\partial \lambda} \Delta(E,\lambda)}{
\frac{\partial}{\partial E} \Delta(E,\lambda)}
\right|_{\substack{\lambda = 0 \\ E = E_m}} \; .
\end{equation}
The corresponding semiclassical approximation is given by
\begin{equation} \label{expectsemi}
\langle \psi_m | A | \psi_m \rangle \approx - \left.
\frac{\Re \left[ \frac{\partial}{\partial \lambda} 
\Delta(E,\lambda) \right]^{\text{sc,tr}}}{
   \Re \left[ \frac{\partial}{\partial E} 
\Delta(E,\lambda) \right]^{\text{sc,tr}}}
\right|_{\substack{\lambda = 0 \\ E = E_m}} \; .
\end{equation}
We insert the semiclassical approximation of
the determinant (\ref{delbassc}) and perform 
the derivatives. This results in 
\begin{equation} \label{expectsemi2}
\langle \psi_m | A | \psi_m \rangle \approx
\frac{\pi \bar{d}_A(E_m) \Im [\Delta(E_m)]^{\text{sc,tr}} 
- \frac{1}{\hbar} B(E_m) \Im \left[ \expp^{- \ci \pi \bar{N}(E_m)}
\sum_{n=1}^\infty A_n \, C_n \exp\left( \frac{\ci}{\hbar} {\cal S}_n \right) 
\right]^{\text{tr}}}{
\pi \bar{d}(E_m)  \Im [\Delta(E_m)]^{\text{sc,tr}}  
- \frac{1}{\hbar} B(E_m) \Im \left[ \expp^{- \ci \pi \bar{N}(E_m)}
\sum_{n=1}^\infty T_n \, C_n \exp\left( \frac{\ci}{\hbar} {\cal S}_n \right) 
\right]^{\text{tr}}} \; .
\end{equation}
Here 
\begin{equation}
\bar{d}_A(E) = \left. \frac{\partial}{\partial \lambda} \bar{N}(E) \right|_{\lambda = 0}
\sim - (2 \pi \hbar)^{-D} \int \ud^D q \, \ud^D p \; 
A(\q,\p) \; \delta(E - H_0(\q,\p)) \, ,
\end{equation}
where $A(\q,\p)$ is the classical function corresponding to the operator $A$.
Furthermore
\begin{equation}
A_n = \sum_{p \in {\cal P}_n} m_p \frac{\partial S_p}{\partial \lambda}
= - \sum_{p \in {\cal P}_n} m_p \int_0^{T_p} A(\q(t),\p(t)) \, \ud t \; ,
\end{equation}
where the integral is evaluated along periodic orbits. 
The last equality can be obtained from \cite{GSBSWZ91,OLM98}.

Note that both, $\bar{d}_A(E)$ and $\bar{d}(E)$, are
proportional to $\hbar^{-D}$. If we take only these
terms into account we obtain the following approximation
\begin{equation}
\langle \psi_m | A | \psi_m \rangle \approx
\frac{\bar{d}_A(E_m)}{\bar{d}(E_m)} \; .
\end{equation}
This is the average of $A(\q,\p)$
over the energy shell, in accordance with quantum ergodicity.

The authors of \cite{EFMW92} express expectation values as a
quotient of derivatives of the zeta function (\ref{zetasemi}) which
is modified by including additional terms in the Euler product.
For the composite orbits these additional terms are factors of the
form $\expp^{-\lambda A_n}$. The resulting approximation differs
from (\ref{expectsemi2}) in that zero-length terms were not
included, and it does not involve a resummation. 
Instead a curvature expansion was applied.
In \cite{KS96} the scattering approach to quantization was
applied to obtain expectation values of smooth operators in
terms of Poincar\'e section wavefunctions.

\section{Discussion}

We have derived in this article semiclassical approximations
for wavefunctions, Green's functions and expectation values 
in terms of a finite number of classical trajectories by
relating them to spectral determinants.
The main results are equations (\ref{psisc}), (\ref{psisc2}),
(\ref{greenresum}), and (\ref{expectsemi}). These formulas 
have a simple and transparent form, and this might facilitate
a theoretical study of semiclassical properties, for example
scars, as well as a numerical evaluation. We have concentrated
on the position representation, but it is not difficult to obtain
other representations as well as is shown by the example of
phase space representations in equations (\ref{wignerresum})
and (\ref{wignerresum2}). Alternatively, one may apply different
perturbations from the start.

One quantity that appears in almost all approximations as an
additional weight of a trajectory $\gamma$ is the semiclassical
spectral determinant which is smoothly truncated at half the Heisenberg
time minus the time along the trajectory, $T_H/2 - T_\gamma$. Hence it
would be of interest to understand the dependence of the truncated
semiclassical spectral determinant on the cut-off. For systems with
zero Maslov index one might investigate this dependence by using
a Guinand-type approach, which involves approximating it by an
integral \cite{Ber91}. 

As is demonstrated by the example in appendix \ref{app_b} the method
can also be applied to one-dimensional systems.
An open question is how the approach of this article is applicable
to integrable systems with dimension $D>1$, or mixed systems.
The relations between wavefunctions, etc., and spectral determinants
in equations (\ref{psi}), (\ref{psi2}), (\ref{delgreen}), and
(\ref{delexpt}) are general and do not depend on the nature of the
classical dynamics. What is needed is an explicit semiclassical
expansion of the spectral determinant in terms of composite orbits
for these systems, or a semiclassical approximation of spectral
determinants that is based on the EBK-quantization for integrable
systems.

\section*{Acknowledgements}

I would like to thank Alfredo Ozorio de Almeida and Raul Vallejos
and other members of the Centro Brasileiro de Pesquisas F\'isicas
for their kind hospitality during a sabbatical leave.
Helpful discussions with Jon Keating, Jens Marklof, Alfredo Ozorio
de Almeida and Jonathan Robbins are gratefully acknowledged.

\appendix

\section{Regularization of spectral determinants} \label{app_a}

This section discusses Weierstrass and zeta regularized spectral
determinants and is based on \cite{Vor87}. It considers
spectral sequences $\{ E_k \}$ with
$0 < E_0 \leq E_1 \leq E_2 \leq \ldots$ and
$E_k \rightarrow \infty$ as $k \rightarrow \infty$.
Further assumptions concern the partition function,
or the trace of the heat kernel. It is assumed that its definition
in terms of a sum over energies is convergent for $\Re t > 0$, and
that it has the following asymptotic expansion
\begin{equation} \label{heat}
\Theta(t) := \sum_{k=0}^\infty \expp^{- t E_k} 
          \sim \sum_{n=0}^\infty c_{i_n} t^{i_n} \; , \quad \text{as}
\quad t \rightarrow 0 \; ,
\end{equation}
with an increasing sequence of real $i_n$ without limit point and $i_0<0$.
Let $\mu = -i_0$ in the following. Equation (\ref{heat}) implies that the
leading term in Weyl's law for the spectral counting function is given by
\begin{equation} \label{weyl}
N(E) \sim \frac{c_{i_0}}{\Gamma(1 - i_0)} E^\mu \; .
\end{equation}
There are different ways to define a spectral determinant by
a convergent product over the eigenvalues $E_k$ of the form
(\ref{delta}). The Weierstrass canonical product is defined by
\begin{equation} \label{weierdet}
\Delta_W(E) = \prod_{k=0}^\infty 
\left( 1 - \frac{E}{E_k} \right) 
\exp \left( \frac{E}{E_k} + 
\frac{E^2}{2 E_k^2} + \dots + 
\frac{E^{[\mu]}}{[\mu] E_k^{[\mu]}} \right) \; ,
\end{equation}
where $[\mu]$ is the integer part of $\mu$.
An alternative regularization, the zeta regularization, 
defines the spectral determinant by an analytic
continuation of the generalized zeta function
\begin{equation} \label{genzeta}
Z(s,a) = \sum_{k=0}^\infty (E_k + a)^{-s} 
= \Gamma(s)^{-1} \int_0^\infty \Theta(t) \, 
\expp^{- a t} \, t^{s-1} \, \ud t \; .
\end{equation}
This definition is valid for $\Re s > \mu$ and $a > -E_0$ and is
extended elsewhere by analytic continuation. The determinant is
defined as
\begin{equation}
\Delta_Z(E) = \exp[-Z'(0,-E)] \; ,
\end{equation}
where the derivative is with respect to the first variable.
Both determinants, $\Delta_W(E)$ and $\Delta_Z(E)$, are entire
functions of $E$. The relation between
both determinants can be understood by looking at the trace of the
$\mu$-th derivative of the resolvent which is given by
the convergent sum
\begin{equation} \label{mugreen}
\Tr G^{([\mu])}(E) = (-1)^{[\mu]} [\mu]! 
\sum_{k=0}^\infty (E - E_k)^{-[\mu]-1} \; .
\end{equation}
(Note that the resolvent is defined with a different sign
in \cite{Vor87}). It can be shown that both determinants can be
obtained by a multiple integration of (\ref{mugreen}). The difference
is that the lower integration limit is $E=0$ for the Weierstrass
regularized determinant whereas it is $E=-\infty$ for the zeta
regularized determinant
\begin{equation} \label{delint}
\log \Delta_W(E) = \left( \int_0^E \right)^{[\mu+1]}
\Tr G^{([\mu])}(E) \; , \qquad \quad
\log \Delta_Z(E) = \left( \int_{-\infty}^E \right)^{[\mu+1]}
\Tr G^{([\mu])}(E) \; .
\end{equation}
In the second case the integral is defined by finite part extractions
when necessary. For example, the following rules hold
\begin{align} \label{finite}
\int_{-\infty}^E (-E')^{-s} \, \ud E' & = 
\frac{(-E)^{1-s}}{s-1} \; , \quad (s \neq 1) \; , \qquad \quad
\int_{-\infty}^E (-E')^{-1} \, \ud E'
= - \log(-E) \; , \qquad
\notag \\
\int_{-\infty}^E \log (-E') \, \ud E' & = 
E (\log(-E) - 1) \; .
\end{align}
It can be further shown that both determinants are related by
\begin{equation} \label{weizetrel}
\Delta_Z(E) = \Delta_W(E) \, \exp\left( - \sum_{m=0}^{[\mu]} a_m E^m \right) \; .
\end{equation}
Expressions for the constants $a_m$ can be found in \cite{Vor87}.

Semiclassical approximations for the spectral determinants are
obtained by starting with the semiclassical expression for the
trace of the $[\mu]$-derivative of the resolvent and doing
the integrations in (\ref{delint}). The zeta regularized determinant
is convenient for semiclassical approximations, because the periodic
orbit terms do not give a contribution at $E = - \infty$. If the
periodic orbit terms are integrated for chaotic systems with unstable
periodic orbits they yield the dynamical zeta function $\zeta(E)$
in (\ref{zetasemi}) \cite{Vor88,BK90}. The asymptotics of $\Tr G^{([\mu]}(E)$
in the limit $E \rightarrow -\infty$ is determined by the zero-length
contributions. From the asymptotic expansion of the trace of the
heat kernel (\ref{heat}) follows that
\begin{equation} \label{gmuasy}
\Tr G^{([\mu])}(E) = - \Gamma([\mu] + 1) \, Z([\mu] + 1, - E) 
\sim - \sum_{n=0}^\infty
c_{i_n} \Gamma([\mu]+1+i_n) (-E)^{-i_n-[\mu]-1} \; ,
\end{equation}
as $E \rightarrow -\infty$. 
For the example of a two-dimensional billiard system
with Dirichlet boundary conditions
(units $\hbar=2 m = 1$) the first three coefficients are
$c_{-1} = A/(4 \pi)$, $c_{-1/2} = -L/(8 \sqrt{\pi})$,
$c_0 = C$, and $\mu=1$. Here $A$ is the area of the
billiard, $L$ is its perimeter and $C$ is determined
by the curvature and corners of the boundary \cite{BH76}.
The integration of (\ref{gmuasy}) with the rules
(\ref{finite}) then yields the zero-length contributions
to the spectral determinant. We state here the result
for two-dimensional billiard systems
\begin{equation} \label{dsemi}
\Delta_Z(E) \approx \exp \left\{ A E (\log(E) - 1)/4\pi + C \log(E)
- i \pi \bar{N}(E) \right\} \zeta(E) \; , 
\end{equation}
where 
\begin{equation}
\bar{N}(E) \approx \frac{A}{4 \pi} E - \frac{L}{4 \pi} \sqrt{E} + C \; .
\end{equation}
Equation (\ref{dsemi}) is in agreement with the appendix of \cite{KS94}.
Comparison with (\ref{delbassc}) shows that in the approximaton (\ref{dsemi})
for the zeta regularized determinant the function $B(E)$ is given by
\begin{equation}
B_Z(E) = \exp \left\{ A E (\log(E) - 1)/4\pi + C \log(E) \right\} \; .
\end{equation}
It is part of the zero-length contribution. For the Weierstrass regularization
$B(E)$ contains an additional factor according to (\ref{weizetrel}).

\section{A one-dimensional example} \label{app_b}

In this section we apply the resummation method to a one-dimensional example.
In order to test the formula for complex-valued Green's functions we consider
a system without time-reversal symmetry. It will be seen that the resummation
is exact if we choose a sharp cut-off by a Heaviside theta function.
\begin{equation}
\theta^{\text{tr}}(T) = \theta(T) = 
\begin{cases} 0 \quad \text{if} \quad T < 0 \\
\frac{1}{2} \quad \text{if} \quad T = 0 \\
1 \quad \text{if} \quad T > 0 \; .
\end{cases}
\end{equation}

The Schr\"odinger equation is given in units where $\hbar = 2 m = 1$ by
\begin{equation} \label{schroe}
\left( - i \frac{\ud}{\ud q} + \alpha \right)^2 \, \psi(q) = E \psi(q) \, ,
\end{equation}
where $\alpha$ is constant, and we apply periodic boundary conditions
of the form $\psi(q+l) = \psi(q)$. The solutions for (\ref{schroe}) are
given by
\begin{equation}
\psi_n(q) = \frac{1}{\sqrt{l}} \expp^{\ci 2 \pi n q/l} \, , \quad
E_n = \left(\frac{2 \pi n}{l} + \alpha\right)^2 \, , \quad n \in \mathbb{Z} \, .
\end{equation}
The corresponding free Green's function is obtained from its
differential equation ($E = k^2$)
\begin{equation}
G^{\text{free}}(q,q',E) = \frac{1}{2 \ci k} \expp^{\ci k |q - q'| - 
\ci \alpha (q - q')} \, .
\end{equation} 
The Green's function of the considered system follows by the method of images
\begin{equation} \label{green1d}
G_0(q,q',E) = \sum_{n = -\infty}^\infty G^{\text{free}}(q + n l,q',E) \, .
\end{equation}
This representation of the Green's function is identical to its
semiclassical approximation. The sum can be interpreted
as sum over all trajectories from $q'$ to $q$ with lengths 
$L_n = |q + n l - q'|$. The sum in (\ref{green1d}) involves
only geometric series and can be carried out with the result
\begin{equation} \label{green1d2}
G_0(q,q',E) = \frac{\expp^{- \ci \alpha (q - q')} \, \left[
\sin(\alpha l) \, \expp^{\ci k (q - q')} + \sin(k l - k |q - q'|)
- \sin(\alpha l - k |q - q'|) \right]}{ 2 k
\left( \cos(\alpha l) - \cos(k l) \right) } \, . 
\end{equation}
For $q'=q$ we obtain
\begin{equation} \label{greendiag1d}
G_0(q,q,E) = \frac{1}{2 \ci k} + \sum_{n=1}^\infty \frac{1}{\ci k}
\expp^{\ci k n l} \, \cos(\alpha n l) = 
\frac{\sin(k l)}{ 2 k
\left( \cos(\alpha l) - \cos(k l) \right) } \, . 
\end{equation}
From the zero-length term we obtain the
mean spectral staircase $\bar{N}(E) = k l / \pi$. 
Instead of the Heisenberg time it is more convenient to speak of
the Heisenberg length which is $L_H = 2 k T_H = 4 k \pi \bar{N'}(E) = 2 l$.
Hence the cut-off for the orbit sums is at $l$.

The spectral determinant is
\begin{equation} \label{delta1d}
\Delta_0(E) = \exp \left( \int \ud E \, \Tr G_0(E) \right) 
= \cos(\alpha l) - \cos(k l) \, .
\end{equation}
This can be written in the form
\begin{equation} \label{delta1d2}
\Delta_0(E) = \expp^{- \ci k l} \left[ - \frac{1}{2} + \cos(\alpha l) 
\expp^{\ci k l} - \frac{1}{2} \expp^{2 \ci k l} \right] \, ,
\end{equation}
which corresponds to the sum over composite orbits in this
one-dimensional system. Applying the resummation with the
theta-function as cut-off we find that
\begin{equation}
\Delta_0(E) \approx 2 \Re [\Delta_0(E)]^{\text{sc,tr}} =
2 \Re \expp^{- \ci k l} \left[ - \frac{1}{2} + 
\frac{1}{2} \cos(\alpha l) \expp^{\ci k l} \right]
= \cos(\alpha l) - \cos(k l) \, .
\end{equation}
Hence the resummation is exact.

Let us apply now the resummation to the Green's function 
\begin{equation}
G_0(q,q',E) \approx \frac{[\Delta_0(E) \, G_0(q,q',E) + 
\overline{\Delta_0(E) \, G_0(q',q,E)} ]^{\text{sc,tr}}
}{2 \Re [\Delta_0(E)]^{\text{sc,tr}}} \; .
\end{equation}
One has to multiply the sums (\ref{green1d}) and (\ref{delta1d2})
and discard all terms for which the joint lengths of the 
trajectories is larger than $L_H/2=l$ (and multiply terms
for which the joint lengths is $l$ by $1/2$). We consider
first the case $q'=q$ and obtain
\begin{align}
G_0(q,q,E) & = 2 \Re \frac{\frac{1}{2 \ci k} \expp^{- \ci k l} 
\left[ -\frac{1}{2} + \frac{1}{2} \cos(\alpha l) \expp^{\ci k l}
- \frac{1}{4} \expp^{\ci k l - \ci \alpha l}
- \frac{1}{4} \expp^{\ci k l + \ci \alpha l}
\right]
}{\cos(\alpha l) - \cos(k l)}
\notag \\ &
= \frac{\sin(k l)}{2 k [\cos(\alpha l) - \cos(k l)]} \; ,
\end{align}
which agrees with (\ref{greendiag1d}). In the case $q' \neq q$ 
\begin{align} \label{greenres1d}
G_0(q,q',E) & = \frac{\frac{1}{4 \ci k} \expp^{- \ci k l} 
\left[ -\expp^{\ci k |q - q'| - \ci \alpha (q - q')} 
- \expp^{\ci k (l - |q - q'|) - \ci \alpha (q - q' \mp l)} \right]
}{\cos(\alpha l) - \cos(k l)} 
\notag \\ & + 
\frac{- \frac{1}{4 \ci k} \expp^{\ci k l} 
\left[ -\expp^{- \ci k |q - q'| - \ci \alpha (q - q')}
- \expp^{- \ci k (l - |q - q'|) - \ci \alpha (q - q' \mp l)} \right]
}{\cos(\alpha l) - \cos(k l)}
\notag \\ & = 
\frac{\sin(k l - k |q - q'|) + \sin(k |q - q'|)\expp^{\pm \ci \alpha l} 
}{2 k [\cos(\alpha l) - \cos(k l)]} \; \expp^{- \ci \alpha (q - q')} \; ,
\end{align}
where the upper and lower signs in the exponents correspond to the
cases $q>q'$ and $q<q'$, respectively. A short calculation shows
that (\ref{greenres1d}) is identical to (\ref{green1d2}).


\end{document}